\documentclass[useASM]{mn2e}
\input{psfig.sty}
\usepackage{graphicx}

\begin{document}

\title{Calibration of Gamma-Ray Burst Luminosity Indicators
\footnote{send offprint request to: Enwei Liang (Email:
lew@physics.unlv.edu)}}

\author[Liang \& Zhang]
       {Enwei Liang$^{1,2}$ and Bing Zhang$^{1}$
 \\
$^1$Physics Department, University of Nevada, Las Vegas,
NV89154; \\lew@physics.unlv.edu, bzhang@physics.unlv.edu\\
$^2$Department of Physics, Guangxi University, Nanning 530004, China }
\maketitle

\label{firstpage}

\begin{abstract}
Several gamma-ray burst (GRB) luminosity indicators have been
proposed, which can be generally written in the form of $\hat {L} =
c\prod x_i^{a_i}$, where $c$ is the coefficient, $x_i$ is the $i$-th
observable, and $a_i$ is its corresponding power-law index. Unlike
Type-Ia supernovae, calibration of GRB luminosity indicators using a
low-redshift sample is difficult. This is because the GRB rate drops
rapidly at low redshifts, and some nearby GRBs may be different
from their cosmological brethren. Calibrating the standard candles
using GRBs in a narrow redshift range ($\Delta z$) near a
fiducial redshift has been proposed recently. Here we elaborate such a
possibility and propose to calibrate $\{a_i\}$ based on the Bayesian
theory and to marginalize the $c$ value over a reasonable range of
cosmological parameters. We take our newly discovered multi-variable
GRB luminosity indicator, $E_{\rm iso}=cE_p^{a_1}t_b^{a_2}$, as an
example and test the validity of this approach through simulations. We
show that while $c$ strongly depends on the cosmological parameters,
neither $a_1$ nor $a_2$ does as long as $\Delta z$ is small
enough. The selection of $\Delta z$ for a particular GRB sample could
be judged according to the size and the observational uncertainty of
the sample. There is no preferable redshift to perform the calibration
of the indices $\{a_i\}$, while a lower redshift is preferable for
$c$-marginalization. The best
strategy would be to collect GRBs within a narrow redshift bin around
a fiducial intermediate redshift (e.g. $z_c\sim 1$ or $z_c\sim 2$),
since the observed GRB redshift distribution is found to peak around
this range. Our simulation suggests that with the current
observational precision of measuring GRB isotropic energy ($E_{\rm
iso}$), spectral break energy ($E_p$), and the optical temporal break
time ($t_{b}$), 25 GRBs within a redshift bin of $\Delta z \sim 0.30$
would give fine calibration to the Liang-Zhang luminosity indicator.
\end{abstract}

\section {Introduction}
The cosmological nature (Metzger et al. 1997) of long gamma-ray bursts (GRBs) and their
association with star formation (e.g., Totani 1997; Paczynski 1998; Bromm \& Loeb 2002)
make GRBs a new probe of cosmology and galaxy evolution (e.g. Djorgovski et al. 2003).
Gamma-ray photons with energy from tens of keV to MeV from GRBs are almost immune to dust
extinction, and should be detectable out to a very high redshift (Lamb \& Reichart 2000;
Ciardi \& Loeb 2000; Gou et al. 2004; Lin et al. 2004). Several plausible GRB luminosity
indicators have been proposed, including luminosity-variability relation (Fenimore \&
Ramirez-Ruiz 2000; Reichart et al. 2001), luminosity-spectral lag relation (Norris,
Marani, \& Bonnell 2000), standard gamma-ray jet energy ($E_{\gamma,jet}$)(Frail et al.
2001; Bloom et al. 2003), isotropic gamma-ray energy ($E_{\rm iso}$) - peak spectral
energy ($E_p$) relation (Amati et al. 2002), $E_{\gamma,jet}-E_p$ relation (Ghirlanda et
al. 2004a), and a multi-variable relation among $E_{\rm iso}$, $E_p$ and the break time
of the optical afterglow light curves ($t_b$) (Liang \& Zhang 2005). Attempts to use
these luminosity indicators to constrain cosmological parameters have been made (e.g.
Schaefer 2003; Bloom et al. 2003; Dai, Liang, \& Xu et al. 2004; Ghirlanda et al. 2004;
Friedmann \& Bloom 2005; Firmani et al. 2005; Liang \& Zhang 2005; Xu, Dai, \& Liang
2005; Xu 2005; Mortsell \& Sollerman 2005; Wang \& Dai 2006). With the discovery of the
tight Ghirlanda-relation (Ghirlanda et al. 2004a) and the more empirical LZ-relation
(Liang \& Zhang 2005), it is now highly expected that GRBs may become a promising
standard candle to extend the traditional Type-Ia SN standard candle to higher redshifts
(e.g. Lamb et al. 2005).

In order to achieve a cosmology-independent standard candle, one needs to calibrate any
luminosity indicator. Otherwise, one inevitably encounters the so-called ``circularity
problem'' (e.g. Firmani et al. 2004; Xu et al. 2005 for discussion). In the case of
Supernova cosmology, calibration is carried out with a sample of Type-Ia SNe at very low
redshift so that the brightnesses of the SNa are essentially independent on the cosmology
parameters (e.g., Phillips 1993; Riess et al. 1995). In the case of GRBs, however, this
is very difficult. The observed long-GRB rate falls off rapidly at low redshifts, as is
expected if long-GRBs follow global star formation. Furthermore, some nearby GRBs may be
intrinsically different. Observations of GRB 980425, GRB 031203 and some other nearby
GRBs indicate that they differ from typical GRBs by showing low isotropic energy, simple
light curve, large spectral lag and dimmer afterglow flux (e.g., Norris 2002; Soderberg
et al. 2004; Guetta et al. 2004; Liang \& Zhang 2006). Although some GRBs in the
optically-dim sample of Liang \& Zhang (2006) still follow the Ghirlanda-relation and LZ-
relation, some very nearby events (e.g. GRB 980425 and GRB 031203) are clearly outliers
of the Amati-relation and also likely the outliers of the Ghirlanda-relation and the
LZ-relation. If at least some low-redshift GRBs are different from their cosmological
brethren, it is very difficult to calibrate the GRB standard candle using a low-redshift
sample.

Recently the possibility of calibrating the standard candles using GRBs in a narrow
redshift range ($\Delta z$) near a fiducial redshift has been proposed (Lamb et al. 2005;
Ghirlanda et al. 2005)\footnote{The similar idea was also discussed in an earlier version
of Liang \& Zhang (2005).}. In this paper we elaborate this method (Lamb et al. 2005;
Ghirlanda et al. 2005) further based on the Bayesian theory. We propose a detailed
procedure to calibrate $\{a_i\}$ with a sample of GRBs in a narrow redshift range
($\Delta z$) without introducing a low-redshift GRB sample and marginalize the $c$ value
over a reasonable range of cosmological parameters. The method is described in \S 2. We
take our newly discovered GRB luminosity indicator as an example to test the approach
through simulations (\S 3). The results are summarized in \S4 with some discussion.

\section{Calibration Method}
A GRB luminosity indicator can be generally written in the form of
\begin{equation}\label{model}
\hat {L}(\Omega)=c(\Omega)Q(\Omega|X; A),
\end{equation}
where $c(\Omega)$ is the coefficient, $\Omega$ is a set of cosmological parameters, and
$Q(\Omega|X; A)$ is a model of the observables $X=\{x_i\}$ (measured in the cosmological
proper rest frame) with the parameter set $A=\{a_i\}$, which is generally written in the
form of $Q(\Omega|X; A)=\prod x_i^{a_i}|_\Omega$. Since in GRB luminosity indicators the
parameters $\{x_i\}$ are usually direct observables (e.g. $E_p$, $t_b$, etc) that only
depend on $z$ but not on the cosmological parameters, the above expression naturally
separates the $\Omega$-dependent part, $c(\Omega)$, from the $\Omega$-insensitive part,
$Q(\Omega|X;A)$. This allows us to develop an approach to partially calibrate the
luminosity indicators without requiring a low-redshift GRB sample. Our approach is based
on the Bayesian theory, which is a method of predicting the future based on what one
knows about the past. Our calibration process can be described as follows.

(1) Calibrate $A$ using a sample of GRBs that satisfy a luminosity indicator and are
distributed in a narrow redshift range $z_0\in z_c\pm \Delta z$. Luminosity distance as a
function of redshift is non-linear, and the dependence of the luminosity distance on the
cosmology model at different redshift is different. Such a sample reduces this non-linear
effect. The parameter set $A$ can be then derived by using a multiple regression method
in a given cosmology $\bar{\Omega}$, $A(\bar{\Omega},z_0)$. The goodness of the
regression is measured by $\chi^2_{\min}(\bar{\Omega},z_0)$,
\begin{equation}\label{chir}
\chi^2_{\min}(\bar{\Omega}, z_0)=\sum_{i}^{N}\frac{[\log
\hat{L}^{i}(\bar{\Omega},z_0)-\log L^{i}(\bar{\Omega},z_0)]^2}{\sigma_{\log
\hat{L}^i(\bar{\Omega},z_0)}^2},
\end{equation}
where $N$ is the size of the sample, $\sigma_{\log \hat{L}^i(\bar{\Omega},z_0)}$ is the
error of the empirical luminosity from the observational errors of observables, and $\log
L^{i}(\bar{\Omega},z_0)$ is the theoretical luminosity. The smaller the reduced
$\chi^2_{\min}(\bar{\Omega},z_0)$, the better the regression, and hence, the higher the
probability that $A(\bar{\Omega},z_0)$ is intrinsic. Assuming that the
$\chi^2_{\min}(\bar{\Omega},z_0)$ follows a normal distribution, the probability can be
calculated by
\begin{equation}\label{weight}
P(\bar{\Omega},z_0)\propto e^{-\chi^2_{\min}(\bar{\Omega},z_0)/2}.
\end{equation}
The calibrated $A$ with a sample distributed around $z_0$ is thus
given by
\begin{equation}
A_0=\frac{\int_{\Omega}A(\bar{\Omega},z_0)P(\bar{\Omega},z_0)
d\bar{\Omega}}{\int_{\Omega}P(\bar{\Omega},z_0)d\bar{\Omega}},
\end{equation}
and its root mean square ($rms$) could be estimated by
\begin{equation}
\delta A_0^2=\frac{\int_{\Omega}[A(\bar{\Omega},z_0)-\bar{A}(z_0)]^2 P(\bar{\Omega},z_0)
d\bar{\Omega}}{\int_{\Omega}P(\bar{\Omega},z_0)d\bar{\Omega}},
\end{equation}
where $\bar{A}_0(z_0)$ is the unweighted mean of $A(\bar{\Omega},z_0)$
in different $\bar{\Omega}$.

(2) Marginalize the $c$ value over a reasonable range for a given GRB sample. The $c$
value depends strongly on the cosmological parameters, so it can only be calibrated with
a low redshift sample. Because of the reasons discussed above, such a low-$z$ sample is
hard to collect. We therefore do not calibrate the $c$ value but rather marginalize it
over a reasonable range of cosmological parameters for a given GRB sample. For a given
$c$ one can derive an empirical luminosity $\hat{L}(\bar{\Omega}, c, A_0, z_0)$ from the
luminosity indicator and its error. The $\chi^2(\bar{\Omega},c, A_0, z_0)$ and the
corresponding probability $P(\bar{\Omega},c, A_0, z_0)$ can be then calculated with the
formulae similar to Eqs. (2) and (3), respectively. Therefore, the calibrated luminosity
is derived by
\begin{equation}
\hat{L_0}=\frac{\int_c\int_{\Omega}\hat{L}(\bar{\Omega},c, A_0, z_0)P(\bar{\Omega}, c,
A_0, z_0)d\bar{\Omega}dc}{\int_c\int_{\Omega}P(\bar{\Omega}, c, A_0,
z_0)d\bar{\Omega}dc},
\end{equation}
and its $rms$ is estimated by
\begin{equation}
\delta \hat{L_0}^2=\frac{\int_c\int_{\Omega}[\hat{L}(\bar{\Omega}, c, A_0,
z_0)-\bar{L}(z_0)]^2 P(\bar{\Omega},c, A_0,
z_0)d\bar{\Omega}dc}{\int_c\int_{\Omega}P(\bar{\Omega},c, A_0, z_0)d\bar{\Omega}dc},
\end{equation}
where $\bar{L}(z_0)$ is the unweighted mean of $\hat{L_0}(\bar{\Omega},c, A_0, z_0)$ in
different $c$ and $\bar{\Omega}$ values.

\section{Simulation Tests}
The current GRB samples that favor various luminosity indicators are very small, so that
one cannot directly utilize our approach to perform the calibration. The calibrations
would nonetheless become possible in the future when enough data are accumulated. We
therefore simulate a large sample of GRBs to examine our approach. The simulations aim to
address the questions such as how many bursts are needed, and how narrow the redshift bin
should be used, etc., given a particular observed sample.  We take the LZ-relation (Liang
\& Zhang 2005) as an example, which reads
\begin{equation}
E_{\gamma, {\rm{iso}}}=c E_{p}^{'a_1}t^{'a_2}_{b},
\end{equation}
where $t^{'}_b$ and $E^{'}_p$ are measured in the cosmic rest frame of the burst proper.
We simulate $10^6$ GRBs. Each simulated GRB is characterized by a set of parameters
denoted by ($z$, $E_p$, $E_{\rm iso}$, $t_b$). It is well known that the $E_p$
distribution of a bright BATSE GRBs presented by Preece et al. (2000) is well modelled by
a Gaussian function. The HETE-2 and Swift observations of X-ray rich GRBs and X-ray
flashes (XRFs; Heise et al. 2000; Lamb, Donaghy, \& Graziani 2005) have considerably
extend the $E_p$ distribution to a softer band. Liang \& Dai (2004) studied the observed
$E_{\rm{p}}$ distribution of GRBs and XRFs, combined with both {\em HETE-2} and BATSE
observations, and found that the observed $E_{\rm{p}}$ distribution for GRBs/XRFs is
bimodal with peaks at $\sim 30$ keV and $\sim 200$ keV. The $\sim 30$ keV peak has a
sharp cutoff at the low energy end, likely being due to the instrument threshold limit. A
recent study of a Swift burst sample marginally reveals such a bimodal distribution
(Zhang et al. 2006). We therefore model the $E_p$ distribution by combining the
observations of BATSE and Swift, i.e.
\begin{eqnarray}
\frac{dp}{d\log E_{p}}&=&\frac{0.70}{0.56\sqrt{\pi/2}}\exp[-2(\frac{\log
E_p-2.30}{0.56})^2]\\
&+&\frac{0.30}{0.56\sqrt{\pi/2}}\exp[-2(\frac{\log E_p-1.55}{0.56})^2]
\end{eqnarray}
with a cutoff at $E_p=30$ keV (see Fig. 1). The $E_{\rm iso}$ distribution is obtained
from the current sample of GRBs with known redshifts. Since the $E_{\rm iso}$
distribution suffers observational bias at the low $E_{\rm iso}$ end, we consider only
those bursts with $E_{\rm iso}>10^{51.5}$ ergs, and get\footnote{Our simulations do not
sensitively depend on the $E_{\rm iso}$ distribution. We have used a random distribution
between $10^{51.5}\sim 10^{54.5}$ ergs, and found that the characteristics of our
simulated GRBs are not significantly changed.} $dp/d\log E_{\rm iso}\propto -0.3 \log
E_{\rm iso}$. The redshift distribution is assumed following the global star forming
history of the universe. The model SF2 of Porciani \& Madau (2001) is used. We truncate
the redshift distribution at 10. A fluence threshold of $S_{\gamma}=10^{-7}$ erg
cm$^{-2}$ is adopted.

We assume that these GRBs satisfy the LZ-relation and derive $t_b$ from the simulated
$E^{'}_p$ and $E_{\rm iso}$. Since the observed $t_b$ is in the range of $0.4\sim 6$
days, we also require that $t_b$ is in the same range to account for the selection effect
to measure an optical lightcurve break. Since the observed $\sigma_x/x$ is about
$10\%-20\%$, the simulated errors of these observables are assigned as $\sigma_x/x=0.25k$
with a lower limit of $\sigma_x/x>5\%$, where $x$ is one of the observables $E_p$,
$S_{\gamma}$, and $t_b$, and $k$ is a random number between $0\sim 1$. Our simulation
procedure is the same as that presented in Liang \& Zhang (2005).

With the simulated GRB sample we examine the plausibility of our calibration approach. We
consider only a flat universe with a varying $\Omega_M$. We picked up two samples with
100 GRBs in each group\footnote{To avoid the statistical fluctuation effect we use a
large sample.}. The first group has a narrow redshift bin (i.e. $z=2.0\pm 0.05$) and the
second group has a wide redshift bin (i.e. $z=2.0\pm 1.0$). We then derive the parameters
$c$, $a_1$, and $a_2$ using the multivariable regression analysis (Liang \& Zhang 2005)
for different cosmological parameter ($\Omega_M$) and evaluate the dependences of the
derived parameters on $\Omega_M$. The dependences of these quantities on $\Omega_M$ are
quantified by the Spearman correlation, and the results are presented in Figure 2. It is
found that $c$ strongly depends on $\Omega_M$ regardless of the value of $\Delta z$, as
is expected. On the other hand, while $a_1$ and $a_2$ are strongly correlated with
$\Omega$ for the case of $\Delta z=1.0$, they are essentially independent of $\Omega_M$
for $\Delta z=0.05$. These results suggest that once $\Delta z$ is small enough the
influence of cosmological parameters on both $a_1$ and $a_2$ becomes significant lower
than the observational uncertainty and the statistical fluctuation. This makes the
calibration of both $a_1$ and $a_2$ possible with a GRB sample within a narrow redshift
bin.

The selection of $\Delta z$ is essential to most optimally establish the calibration
sample. Two effects are needed to take into consideration to select $\Delta z$, i.e. the
observational errors of the sample and the statistical fluctuation effect. The most
optimal calibration sample requires that the variations of the standard-candle parameters
caused by varying cosmology should be comparable to the variations caused by these two
effects. In such a case we could establish a sample with a large enough $N$ to reduce the
fluctuation effect while in the mean time with a small enough $\Delta z$ so that the
dependences of both $a_1$ and $a_2$ on $\Omega_M$ are not dominant. Since the relation
between $a_1$ (or $a_2$) and $\log \Omega_M$ is roughly fitted by a linear function (see
Figure 2), we measure the dependence by the chance probability ($P$) of the Spearman
correlation. If $P<10^{-4}$ the dependence is statistical significant, and the sample is
inappropriate for the calibration purpose. Figure 3 shows the distributions of $\log P$
for $a_1$ (left) and $a_2$ (right) in the $\Delta z$-$N$ plane, assuming the current
observational errors for the observables. The grey contours mark the areas that the
dependences of $a_1$ and $a_2$ on $\Omega_M$ are statistical significant. We find that
$P$ dramatically decreases as $\Delta z$ increases for a given $N$. Given a $P$ value,
$\Delta z$ initially decreases rapidly as $N$ increases but flattens at $N>50$. This
indicates that the statistical fluctuation effect is much lower than the observational
errors for a sample with $N>50$. We can see that with the current observational
precision, $\Delta z\sim 0.3$ is robust enough to calibrate both $a_1$ and $a_2$.
Increasing the GRB sample size alone does not improve the calibration when $N>50$, since
the $a_1$ and $a_2$ errors are dominated by the observational uncertainties in the data.
In order to improve calibration further, higher observational precision of $E_{iso}$,
$E_p$ and $t_b$ is needed, which requires a broad-band $\gamma$-ray detector and good
temporal coverage of the afterglow observations.

The observed GRB redshift distribution ranges from 0.0085 to 6.3. We examine if there
exists a preferable redshift range for the calibration purpose. We randomly select a
sample of $25$ GRBs at $z_c\pm 0.3$, and perform the multivariable regression analysis to
derived $a_1$ and $a_2$ from this sample by assuming a flat universe with
$\Omega_M=0.28$. The derived $a_1$ and $a_2$ are plotted as a function of $z_c$ in Figure
4. We find that they are not correlated with $z_c$, and their variations are essentially
unchanged, i.e. $\sim 0.15$. This indicates that there is no evidence for a vantage
redshift range to calibrate $a_1$ and $a_2$. It is therefore equivalent to select a
sample at any redshift bin to calibrate $a_1$ and $a_2$. Such a sample is likely to be
established with GRBs at $z_c=(1-2.5)$, since the observed redshift distribution peaks in
this range. The cosmological dependence is less significant at lower redshifts. So, a
lower redshift (e.g. $z_c=1$) sample is preferred for $c$ marginalization.

According to Figure 3, the best strategy to perform GRB standard candle calibration is to
establish a moderate GRB sample (e.g. 25 bursts) within a redshift bin of $\Delta z \sim
0.3$ at a fiducial intermediate redshift (e.g. $z_c \sim 1$ or $z_c \sim 2$). We simulate
a sample of GRB with $N=25$, $z=1\pm 0.3$, and derive $a_1$ and $a_2$ as a function of
$\Omega_M$ in Figure 5. The calibrated $a_1$ and $a_2$ are $1.93\pm 0.07$ and $-1.23\pm
0.07$, respectively, where the quoted errors are at $3\sigma$ significance level.

\begin{figure}
\begin{center}
\includegraphics[width=3.in,angle=0]{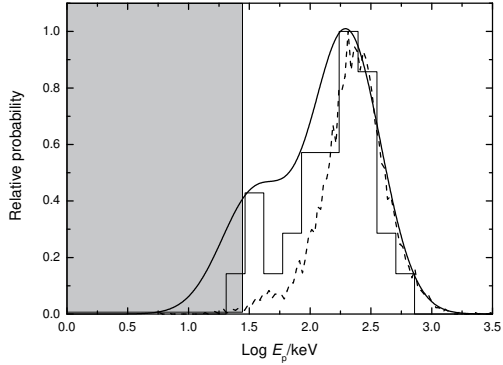}
\end{center}
\caption{The observed $E_p$ distribution: {\em dashed line} ---
derived from bright GRB sample (Preece et al. 2000); {\em step-line}
--- Swift data (Zhang et al. 2006); {\em smoothed-curve} --- our model
with bimodal Gaussian distribution (Eq. 9). The dark region marks the
cutoff at $E_p<30$ keV due to the instrument threshold limit.}
\end{figure}

\begin{figure}
\begin{center}
\includegraphics[width=3.in,angle=0]{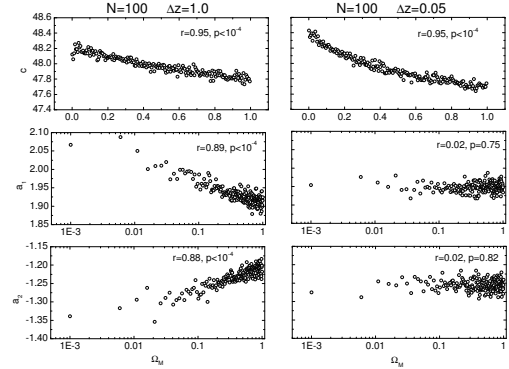}
\end{center}
\caption{Comparison of the dependences of $c$, $a_1$, and $a_2$ on
$\Omega_M$ for a sample of 100 GRBs distributed in $z=2.0\pm 1.0$
(left panels) and in $z=2.00\pm 0.05$ (right panels),
respectively. Current observational errors are introduced for the
simulated bursts. The dependences are measured by the Spearman
correlation, and the correlation coefficient ($r$) and its chance
probability ($P$) are marked in each panel.}
\end{figure}

\begin{figure}
\begin{center}
\includegraphics[width=3.in,angle=0]{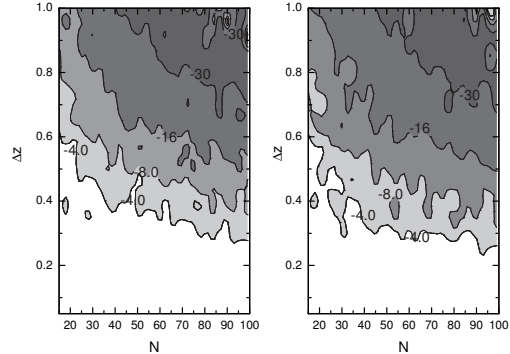}
\end{center}
\caption{Distribution of $\log P$ in the ($N$, $\Delta z$)-plane. The
grey contours mark the areas where the dependences of $a_1$ and $a_2$
on $\Omega_M$ are statistical significant ($P<10^{-4}$). The white
region is suitable for the calibration purpose.}
\end{figure}

\begin{figure}
\begin{center}
\includegraphics[width=3.in,angle=0]{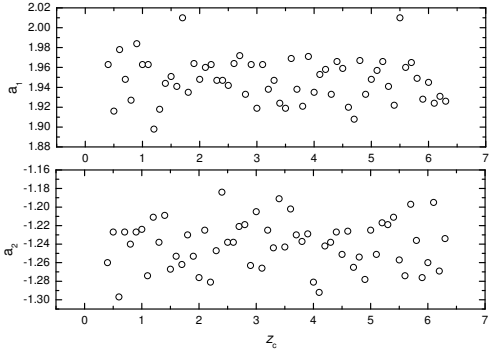}
\end{center}
\caption{The variations of $a_1$ and $a_2$ as a function of $z_c$. The calibration sample
consists of 25 simulated GRBs at $z_c\pm 0.3$. $\Omega_M = 0.28$ is adopted.}
\end{figure}

\begin{figure}
\begin{center}
\includegraphics[width=3.in,angle=0]{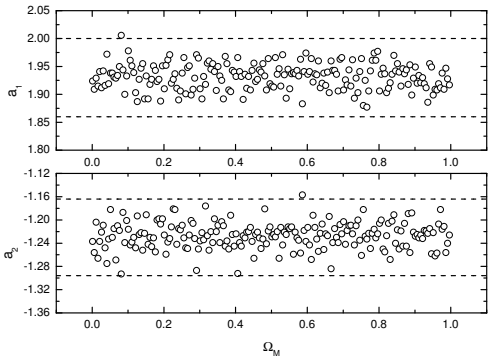}
\end{center}
\caption{The variations of $a_1$ and $a_2$ as a function of $\Omega_M$. The calibration
sample consists of 25 simulated GRBs distributed in the redshift bin $z=1\pm 0.3$. The
dashed lines enclose the $3\sigma$ significance regions.}
\end{figure}

\section{Conclusions and Discussion}
We have explored in detail an approach to calibrate the GRB luminosity indicators, $\hat
{L}(\Omega)=c(\Omega)Q(\Omega|X; A)$, based on the Bayesian theory without a low redshift
GRB sample. The essential points of our approach include, (1) calibrate $A$ with a sample
of GRBs in a narrow redshift bin $\Delta z$; and (2) marginalize the $c$ value over a
reasonable range of cosmological parameters for a given GRB sample. We take our newly
discovered multi-variable GRB luminosity indicator $E_{\rm iso}=cE_p^{a_1}t_b^{a_2}$
(LZ-relation) as an example to test the above approach through simulations. We show that
while $c$ strongly depends on cosmological parameters, both $a_1$ and $a_2$ do not if
$\Delta z$ is small enough. The selection of $\Delta z$ depends on the size and the
observational uncertainty of the sample. For the current observational precision, we find
$\Delta z \sim 0.3$ is adequate to perform the calibration.

It is also found that the calibrations for both $a_1$ and $a_2$ are equivalent for
samples at any redshift bin. The best strategy would be to collect GRBs within a narrow
redshift bin around a fiducial intermediate redshift (e.g. $z_c\sim 1$ or $z_c\sim 2$),
since the observed GRB redshift distribution is found to peak in this range. Our
simulation suggests that with the current observational precision of measuring GRB
isotropic energy ($E_{\rm iso}$), spectral break energy ($E_p$), and the optical temporal
break time ($t_{b}$), 25 GRBs within a redshift bin of $\Delta z \sim 0.30$ would give
fine calibrations to the LZ-relation. Inspecting the current GRB sample that satisfies
the LZ-relation, we find that nine GRBs, i.e. 970828 ($z=0.9578$), 980703 ($z=0.966$),
990705 ($z=0.8424$), 991216 ($z=1.02$), 020405 ($z=0.69$), 020813 ($z=1.25$), 021211
($z=1.006$), 041006 ($z=0.716$), and 050408 ($z=1.24$) are roughly distributed in the
redshift range $z=1.0\pm 0.3$. We expect roughly 15 more bursts to form an adequate
sample to calibrate the LZ-relation.

The observed redshift distribution for the current long GRB sample covers from 0.0085 to
6.29. There have been suggestions that GRB properties may evolve with redshift (e.g.
Lloyd-Ronning et al. 2002; Amati et al. 2002; Wei \& Gao 2003; Graziani et al. 2004;
Yonetoku et al. 2004). Among the proposed GRB luminosity indicators the cosmological
evolution effect has not been considered. With the current GRB sample with known
redshifts, it is difficult to access whether and how GRBs evolve with redshift.
Nonetheless, since our calibration approach makes use of a GRB sample in a narrow
redshift bin, the evolution effect essentially does not affect on the calibration of the
parameter set $A$, the set of the power index (indices) in the luminosity indicators.
However, it could significantly impact on the $c$-marginalization.

We appreciate the valuable comments from the referees. We also thank Z. G. Dai for
helpful discussion and G. Ghirlanda for constructive comments.  This work is supported by
NASA under NNG05GB67G, NNG05GH92G, and NNG05GH91G (BZ \& EWL), and the National Natural
Science Foundation of China (No. 10463001, EWL).

\end{document}